# Installation of the Dark Energy Spectroscopic Instrument at the Mayall 4-meter telescope

Robert Besuner[*a], Lori Allen[b], Charles Baltay[c], David Brooks[d], Pierre-Henri Carton[e], Andrew Peter Doel[d], John Donaldson[b], Yutong Duan[f], Patrick Dunlop[b], Jerry Edelstein[a], Matt Evatt[b], Parker Fagrelius[b], Enrique Gaztañaga[g], Derek Guenther[b], Gaston Gutierrez[h], Michael Hawes[b], Klaus Honscheid[i], Pat Jelinsky[a], Richard Joyce[b], Armin Karcher[j], Martin Landriau[j], Michael Levi[j], Christophe Magneville[e], Robert Marshall[b], Paul Martini[i], Daniel Pappalardo[i], Claire Poppett[a], Francisco Prada[k], Ashley J. Ross[i], Michael Schubnell[l], Ray Sharples[m], William Shourt[a], Joseph Harry Silber[j], David Sprayberry[b], Bob Stupak[b], Gregory Tarle[l], Kai Zhang[j]

[a]Space Sciences Laboratory, University of California, Berkeley, 7 Gauss Way, Berkeley, CA 94720, USA; [b]NSF's National Optical-Infrared Astronomy Research Laboratory, Tucson, AZ; [c]Yale University, New Haven, CT; [d]University College London, UK; [e]Commissariat à l'énergie atomique (CEA)-Saclay, France; [f]Boston University, Boston, MA; [g]Institute of Space Sciences (ICE, CSIC), Barcelona, Spain; [h]Fermi National Accelerator Laboratory, Batavia, IL; [i]Department of Physics, The Ohio State University, Columbus, OH; [j]Lawrence Berkeley National Laboratory, CA; [k]Instituto de Astrofisica de Andalucia CSIC, Granada, Spain; [l]University of Michigan, Ann Arbor, Michigan; [m]Durham University, UK

## ABSTRACT

The Dark Energy Spectroscopic Instrument (DESI) is a Stage IV ground-based dark energy experiment that will measure the expansion history of the Universe using the Baryon Acoustic Oscillation technique. The spectra of 35 million galaxies and quasars over 14000 square degrees will be measured during the life of the experiment. We describe the installation of the major elements of the instrument at the Mayall 4m telescope, completed in late 2019. The previous prime focus corrector, spider vanes, and upper rings were removed from the Mayall's Serrurier truss and replaced with the newly-constructed DESI ring, vanes, cage, hexapod, and optical corrector. The new corrector was optically aligned with the primary mirror using a laser tracker system. The DESI focal plane system was integrated to the corrector, with each of its ten 500-fiber-positioner petal segments installed using custom installation hardware and the laser tracker. Ten DESI spectrographs with 30 cryostats were installed in a newly assembled clean room in the Large Coude Room. The ten cables carrying 5000 optical fibers from the positioners in the focal plane were routed down the telescope through cable wraps at the declination and hour angle axes, and their integral slitheads were integrated with the ten spectrographs. The fiber view camera assembly was installed to the Mayall's primary mirror cell. Servers for the instrument control system replaced existing computer equipment. The fully integrated instrument has been commissioned and is ready to start its operations phase.

**Keywords:** BAO, DESI, baryon, Multi-object spectrograph, laser tracker, Mayall, dark energy,

## 1. INTRODUCTION

The U.S. Department of Energy's Dark Energy Spectroscopic Instrument (DESI) studies Dark Energy by measuring the expansion history of the Universe to unprecedented precision[1]. It employs the Baryon Acoustic Oscillation (BAO) technique, in which the density distribution of baryons (ordinary matter) in the Universe is mapped by way of obtaining redshifts of a large sample of galaxies. DESI will take spectra of 35 million galaxies and quasars over 14,000 square degrees during its planned five year survey duration. DESI was constructed with Lawrence Berkeley National Laboratory (Berkeley Lab) as the project lead laboratory and is now installed at the Mayall 4-meter telescope on Kitt Peak, AZ, a facility operated by NSF's National Optical-Infrared Astronomy Research Laboratory (NOIRLab, formerly NOAO).

---

[*] rwbesuner@lbl.gov; phone 1 (510) 486-5698

DESI is a complex instrument, consisting of several major subsystems, tightly integrated together and with the Mayall telescope and building. The DESI installation process required extensive, detailed planning and coordination by numerous institutions, which began in 2011. From 2016 to 2019, the NOIRLab staff and the constructors of the DESI hardware successfully installed all the components of DESI at the Mayall, transforming the 50-year-old facility into a cutting-edge cosmology research tool.

While some DESI installation activities had begun in 2016, the Mayall continued to operate as an astronomy community user facility until it was shut down on February 12, 2018 in order to begin facility preparations for installation of the major DESI systems.

## 2. INSTRUMENT CONTROL SYSTEM

The DESI Instrument Control System (ICS) communicates with the Mayall telescope and controls and monitors all the DESI elements[2]. Its development was lead by DESI personnel at The Ohio State University (OSU). It consists of software installed on racks of computers located in the climate-controlled computer room on the M floor (main floor, which is the bottom of the dome enclosure volume) of the Mayall building. The system can be operated remotely or on-site. The first elements of the ICS software were integrated with the Mayall Telescope Control System[3,4] in 2016 (for operation of ProtoDESI, an early technology demonstrator[5]), with the full production software package and DESI-specific computing hardware (Figure 1, left) installed on the mountain in May 2018.

As part of the Mayall preparations for the DESI era, an area of the building's U floor (utility floor, two levels below the M floor) was converted by NOIRLab staff from a general-use space to a well-equipped control room[6] (Figure 1, right). This space comfortably accommodates the large number of personnel required for DESI commissioning and operations, which would have been impossible in the existing control room on the C floor (console floor, two levels above the M floor). Nevertheless, the C floor control room retains its functionality, being especially useful for operating the telescope and dome for daytime maintenance work, due to its proximity to and direct view of the telescope.

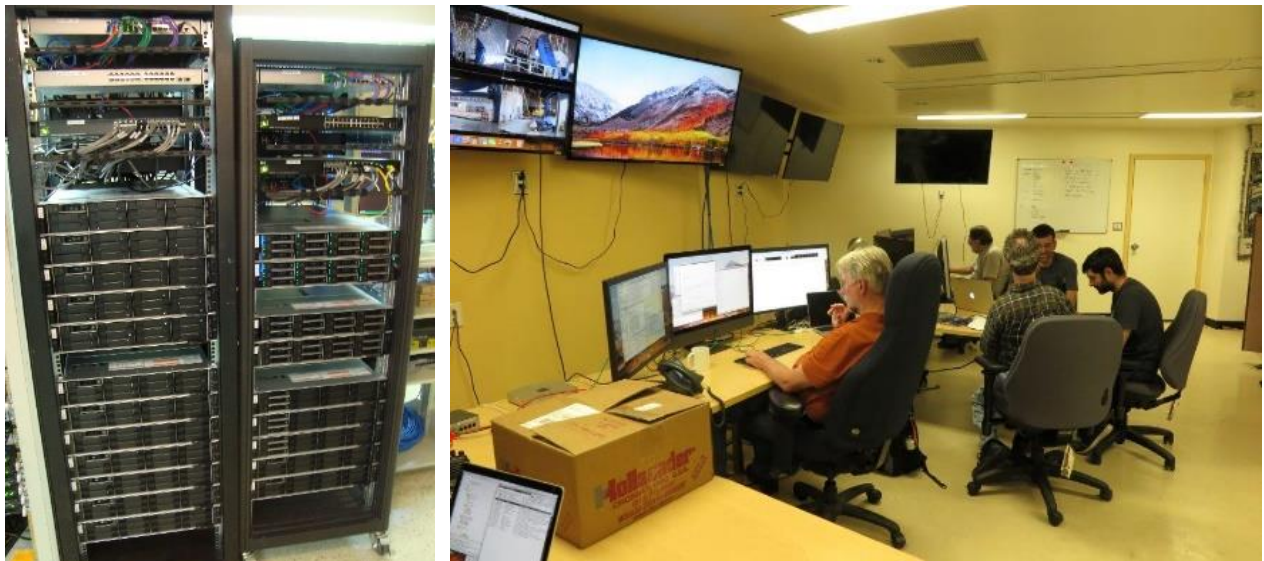

Figure 1: Left-DESI ICS computer racks in the M floor computer room. Right-DESI and NOIRLab personnel operate and monitor the system from the comfort of the Mayall's U floor control room.

## 3. UPPER ASSEMBLY AND CORRECTOR

The DESI upper assembly, designed by Fermi National Accelerator Laboratory (Fermilab), supports the DESI corrector and focal plane assembly atop the Mayall's Serrurier truss[7]. To facilitate the wide field of view required for the DESI survey, the project constructed a new six-lens prime focus corrector, with design of the lens cells and integration of the optics conducted by University College London (UCL) in the United Kingdom[8]. The corrector is comprised of three steel barrel segments from Fermilab, each supporting two lenses on lens cells, with all three corrector segments precisely aligned and pinned with each other for repeatable disassembly and assembly. The corrector is supported by the precision hexapod (a Fermilab responsibility), providing six degrees of freedom of adjustment for corrector focus and alignment relative to the Mayall's primary mirror. The hexapod is supported by the welded steel cage assembly, which is supported from the new DESI upper ring by twelve adjustable-length spider vanes. The old Mayall corrector support system included two rings, one ring fixed to the truss, supporting the pivots and drives for the other ring, which supported the cage and rotated 180 degrees to switch between prime focus and Cassegrain instruments. This capability is not required for DESI, so that assembly was replaced with a single ring fixed to the truss, helping to accommodate the increased mass of the DESI cage, hexapod, corrector, and focal plane assembly as compared to the old corrector, secondary mirror and imaging optics/instrument. The total mass of the DESI hardware supported by the telescope truss was engineered to match the 10,700 kg of existing hardware removed. The DESI upper assembly and corrector systems were integrated and aligned at the Mayall between June 2018 and February 2019.

### 3.1 Removal of existing upper assembly

NOIRLab hired a 300 foot telescoping mobile crane to remove the old Mayall upper assembly intact from the telescope (Figure 2). In preparation for this lift, existing air, data, and electrical services to the ring were disconnected or severed, and ancillary or readily detachable hardware was removed. The four pairs of support struts that make up the Serrurier truss would be unstable with the upper ring removed, so Kitt Peak staff constructed a two-level work platform and ladder system that both provided personnel access to the upper assembly when parked at zenith and braced the Serrurier truss elements while no upper ring was present (Figure 3, left). Kitt Peak designed and built temporary hydraulic jack assemblies that attached to the tops of the Serrurier truss whose purpose was to apply lifting forces on the old upper ring with fine control to ensure safe and controlled separation of the ring. These were a mitigation against difficulties removing the steel alignment pins between the ring and truss, which had not been disturbed since they were originally installed around 50 years before. The jacks were used to slightly raise the ring off the truss pads, ensuring that when the crane lifted the ring, the crane load was only the dead weight of the old upper assembly.

The crane lifted the assembly from the top of the telescope, through the open dome slit, to a flatbed truck that removed it for storage. While the Mayall dome has an integral 50-ton crane, using the external crane posed less risk and schedule impact on the DESI installation process, largely due to the fact that the large DESI elements were already being staged within the Mayall building, with space at a premium. Landing, dismantling, and extracting the old upper assembly inside the building would have interfered substantially with the process of integrating the new DESI upper assembly, slowing DESI installation.

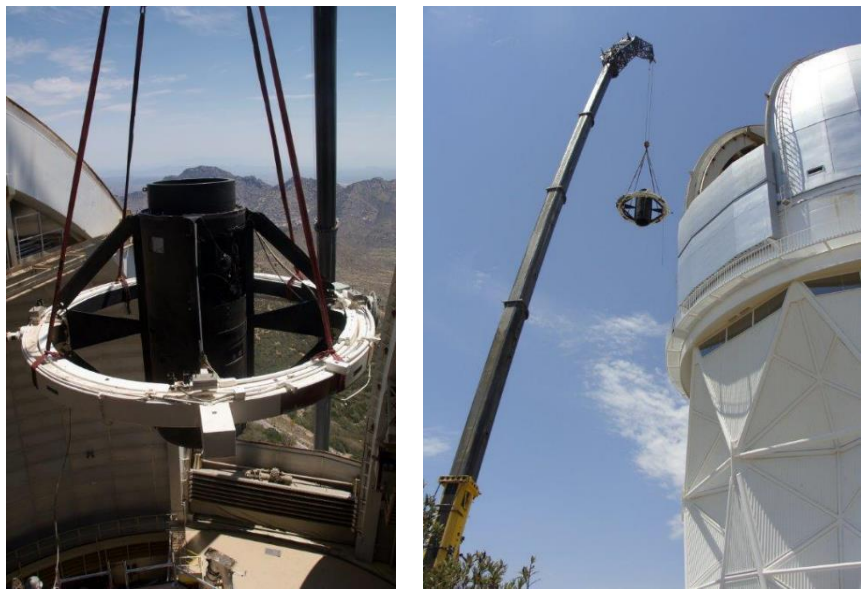

Figure 2: Left-The old Mayall prime focus cage, support vanes, and support rings being lifted away from the telescope through the open dome slit. Visible through the dome slit are part of the crane boom, the nearby Coyote Mountains, and the more distant Tucson Mountains, behind which lies the city of Tucson.  Right-The old upper assembly is swung clear of the dome and lowered to a waiting flatbed truck.

### 3.2 Alignment and match drilling of upper ring

With the old Mayall upper assembly removed, the DESI upper ring was hoisted to the top of the Serrurier truss, where it was aligned to the primary mirror, match drilled with the truss pads for alignment pins, and marked for the locations of the bolts that fasten it to the truss.  Although there is a complete set of fabrication prints for the existing telescope structure, match drilling for pins and transferring bolt hole locations from the truss to the ring guaranteed proper fit of the new DESI ring.

In its final assembled and installed configuration, the DESI cage is supported by the upper ring via twelve tensioned, adjustable spider vanes.  In order to maintain torsional stiffness between the ring and the cage, these vanes are angled 11 degrees relative to radial such that they are able to carry a tangential component of force between the ring and cage.  For the alignment and transfer of mounting features, the upper ring was put into a similar state of stress and deformation as it would when installed with the cage.  This was accomplished by applying tension using the four DESI middle vanes connected to an assembly built by NOIRLab that replicated the vane attachment geometry of the cage, deforming the ring from its unloaded shape by a few millimeters (Figure 3, right).

The pre-tensioned upper ring was lifted to the top of the Serrurier truss using the 5-ton dome crane (Figure 3, left), where it was centered relative to the primary mirror using an alignment telescope in the mirror central hole.  Once centered, holes for the new 5/8" diameter alignment pins were match-drilled, and the locations on the ring were marked to tap for the ¾-10 bolts that fasten the ring to the truss.  The ring was then brought back to the floor and flipped over for drilling and tapping (Figure 3, right).

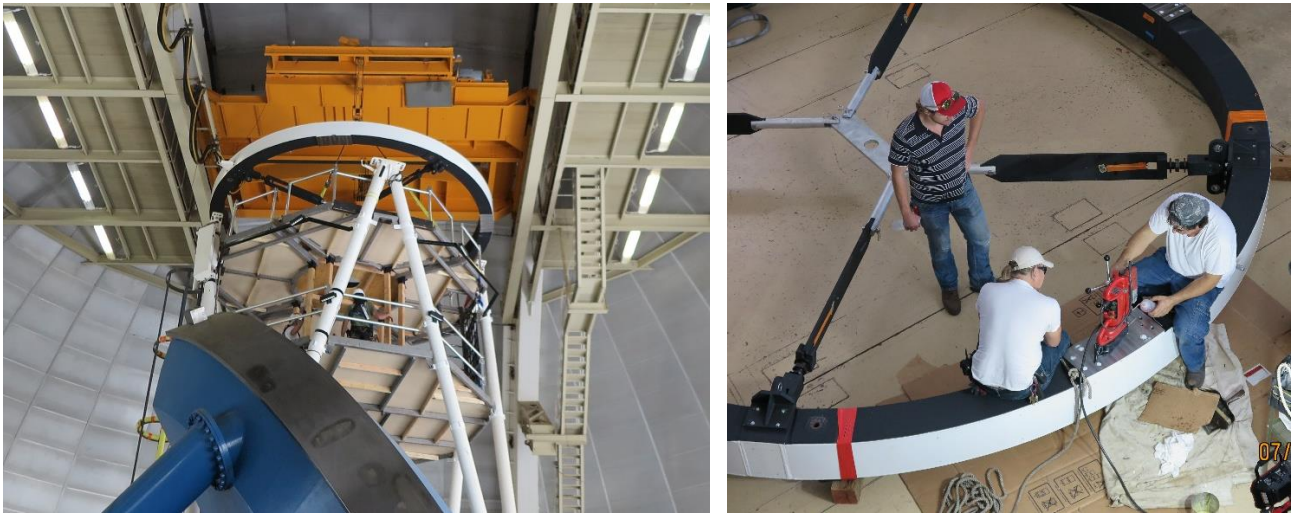

Figure 3:  Left-The pre-tensioned DESI upper ring being lowered by the 5-ton dome crane onto the top of the Serrurier truss for alignment, match-drilling, and transfering the locations where holes are to be tapped into the ring for its mounting bolts.  The four pairs of struts comprising the Serrurier truss are held in place by the two-level, temporary work platforms, which were installed prior to removal of the old upper assembly.  Right-NOIRLab personnel field-tapping the bottom side of the DESI upper ring at the locations marked during the operation shown in the left photo.  The pre-tensioning system, consisting of the four DESI middle vanes and a custom spreader assembly in the center, remains in place and tensioned, to allow test-fitting of the ring and fasteners to the Serrurier truss.

### 3.3 Re-assembly of corrector and integration with hexapod

The three corrector segments were shipped disassembled from the U.K. to Tucson, Arizona on a project-chartered cargo airplane to minimize unsupervised transfers and handling that would have occurred if they were shipped on scheduled flights.  In the same shipment were the alignment spin table and other equipment used by UCL to align the optics in the U.K.  This equipment was set up on the ground floor of the Mayall building and was used to re-assemble the corrector and re-verify its alignment post-shipping (Figure 4)[9].  This verification used the rotating precision table to measure the mechanical runout of the corrector elements and to perform "pencil-beam" tests, in which a laser was aimed along the corrector axis and reflected off a mirror back through the optics, where its wobble was measured on a CCD sensor to confirm lens alignment. These tests replicated the results from the same tests pre-shipment in the U.K., confirming that alignment was not affected during shipment.

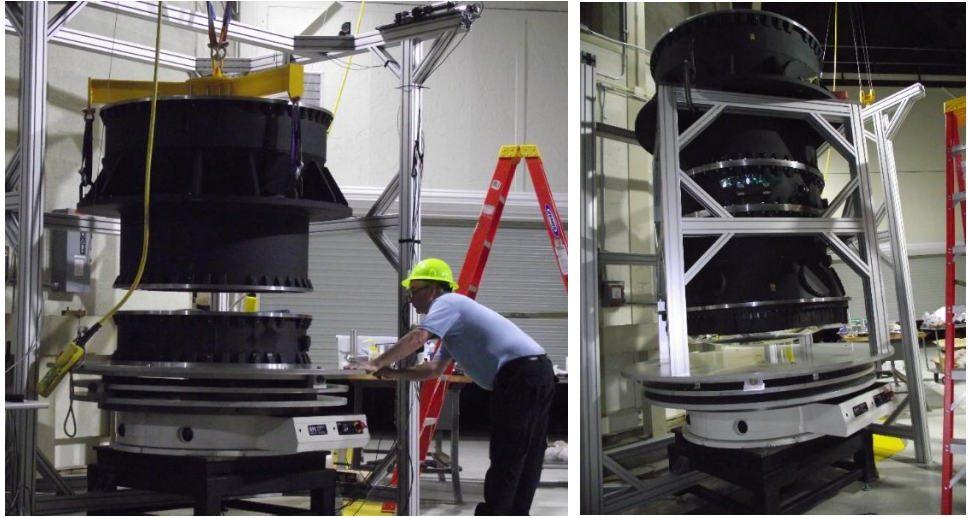

Figure 4: Left-The DESI corrector aft and middle sections being re-assembled by UCL's Dr. David Brooks at the Mayall on the UCL precision turntable for post-shipment alignment verification. Right-The fully re-assembled corrector on the alignment turntable.

After post-shipment verification, the corrector was partially disassembled, with the forward section separated from the other two, to allow installation of the hexapod on the integration stand (Figure 5, left). The diameters of the forward-most and rearward-most corrector lenses are larger than the lenses between. The hexapod inside diameter is smaller than either end of the corrector barrel, so the hexapod is captive to the barrel when fully assembled.

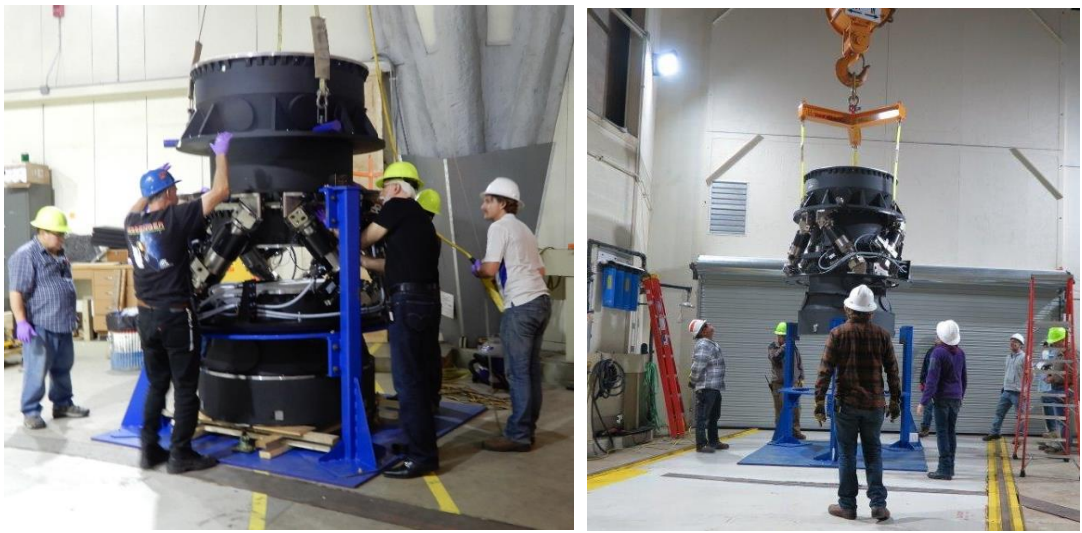

Figure 5: Left-The corrector aft and middle sections being lowered by crane through the hexapod assembly onto the forward section, sitting on wooden blocking in the blue integration frame. Right-The fully-assembled corrector and hexapod being lifted by the Mayall dome crane from the integration stand on the ground floor, on its way to the cage/vanes/ring assembly on the C floor in the Mayall dome.

The dome crane was then used to hoist the combined assembly of the corrector and hexapod from the integration stand on the ground floor of the Mayall building (Figure 5, right), up through the open hatch of the dome floor, and into the cage of the corrector support structure whose preparations are described in the next subsection.

## 3.4 Assembly of cage, vanes, and upper ring

After preparing the ring for attachment to the Mayall Serrurier truss, it was flipped back upright and placed on tall jack stands on the C floor, just east of the telescope itself. The cage was lowered onto blocks roughly centered in the ring, and the twelve spider vanes were installed loosely, connecting the cage with the ring. The cage was then centered and its axis aligned with that of the ring by adjusting the lengths of the vanes and measuring the cage and ring positions and orientations using a laser tracker (Figure 6, left). At this time, the vanes were left loose so the top ring of the cage could be removed without any preload in it. The top ring of the cage was removed (Figure 6, right) to allow later insertion of the corrector and hexapod from above.

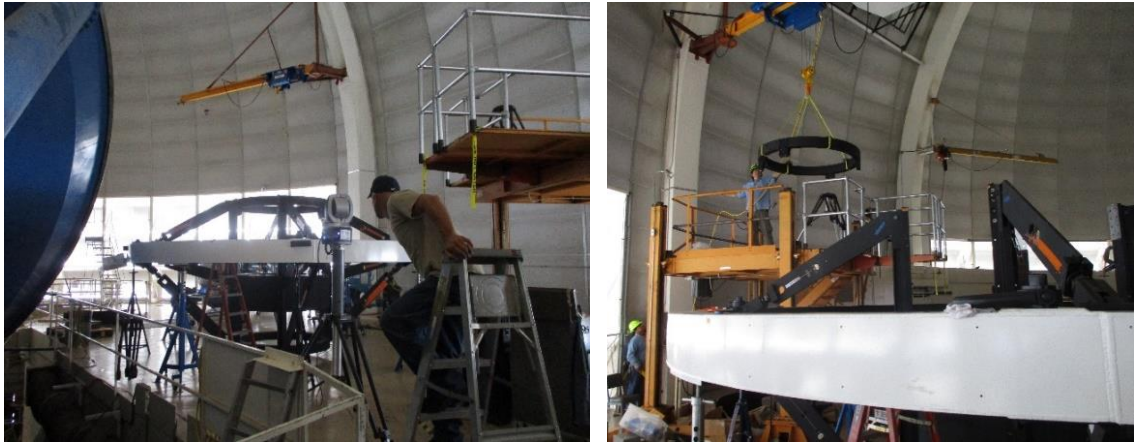

Figure 6: Left-The DESI cage being centered relative to the upper ring on the Mayall C floor. The spider vanes' lengths were adjusted to move and orient the cage, with positioning feedback from the laser tracker in the foreground. Right-The top ring of the cage being lowered onto the yellow Southeast Platform, after temporary removal to allow installation of the corrector and hexapod into the cage.

## 3.5 Assembly of the corrector/hexapod with the cage, spider vanes, and upper ring

The corrector and hexapod were lifted from the ground floor of the Mayall and lowered into the cage/ring assembly on the C floor (Figure 7). The lower flange of the hexapod was bolted to the cage, and the cage top ring was re-installed, completing the cage structure (Figure 8). In preparation for installation to the telescope, the vanes were adjusted to pre-determined tensions, such that they all remain in tension regardless of in-service telescope orientation. This helps maintain alignment accuracy by avoiding uncertain vane length when vane loads vary between tension and compression.

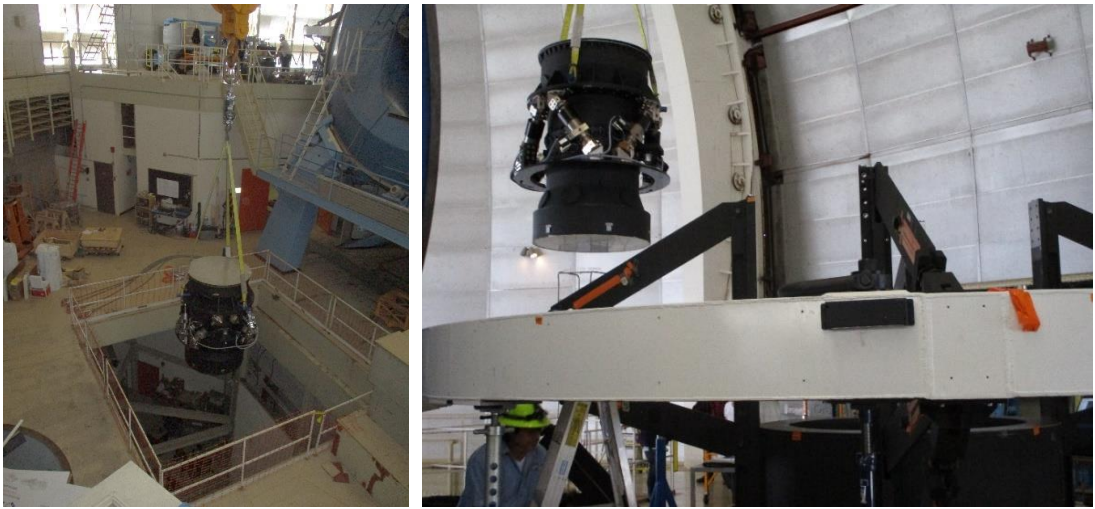

Figure 7: Left-The DESI corrector and hexapod being lifted from the ground floor through the open hatch in the M floor, just north of the blue telescope mount. The ring and vane assembly is visible near the top right of the photo on the C floor,

mostly obscured by the telescope's "horseshoe". Right-The corrector/hexapod assembly nearing the end of its journey by crane from the integration stand on the ground floor to the pre-aligned corrector support structure. The top ring of the cage was temporarily removed to clear the hexapod flange for this operation.

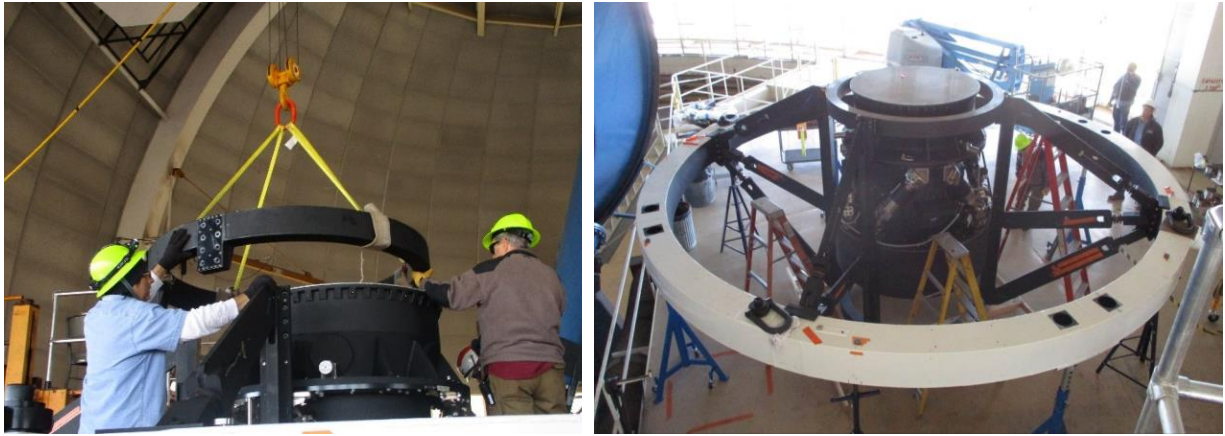

Figure 8: Left-The cage top ring is reinstalled after the corrector and hexapod installation to the cage. Right-The corrector/hexapod installed in the cage, with the cage upper ring reinstalled, the corrector centered relative to the upper ring, and the vanes adjusted to their in-service tensions.

Each spider vane features a screw with left-handed threads on one end and right-handed threads on the other, enabling its length to be controlled by turning that screw. Each vane has strain gages bonded to both sides of the flat blade section, which enabled real-time monitoring of all the vane tensions during the process of preloading the system. These strain gages were calibrated by hanging dead weights from the vanes in a vertical configuration. Because there are twelve spider vanes controlling six degrees of freedom of position and orientation of the cage with respect to the ring, it is a statically indeterminate system. The process of tensioning the vanes was iterative, aided by a spreadsheet tool, based on finite element analyses, that predicted what changes in force of all the vanes would result from a change in one vane's length. To maintain centration of the cage, the upper four, middle four, and lower four sets of vanes were shortened similar amounts by their vertical position in the assembly. After tensioning was completed, the laser tracker was used to verify the position and alignment of the cage were maintained within spec.

The upper assembly was integrated to the top of the telescope without the focal plane assembly installed. DESI was designed to maintain the telescope balance in-service, with the focal plane installed. Because integration of the focal plane assembly was performed at the Southeast Platform, the telescope had to be balanced for it to slew there from zenith. Balance weights were clamped to the ring while it was easily accessible on the jack stands on the C-floor, prior to lifting the assembly onto the telescope.

### 3.6 Installation and alignment of the DESI upper assembly

The Mayall 50-ton dome crane lifted the 10,700 kg DESI upper assembly to the top of the Serrurier truss with guidance from Kitt Peak engineers in a personnel lift (Figure 9, left). Once it was in place, NOIRLab staff on work platforms installed the alignment pins and bolts to secure the ring to the truss (Figure 9, right). A laser tracker mounted on the telescope center section was then used to help align the corrector with the Mayall primary mirror.

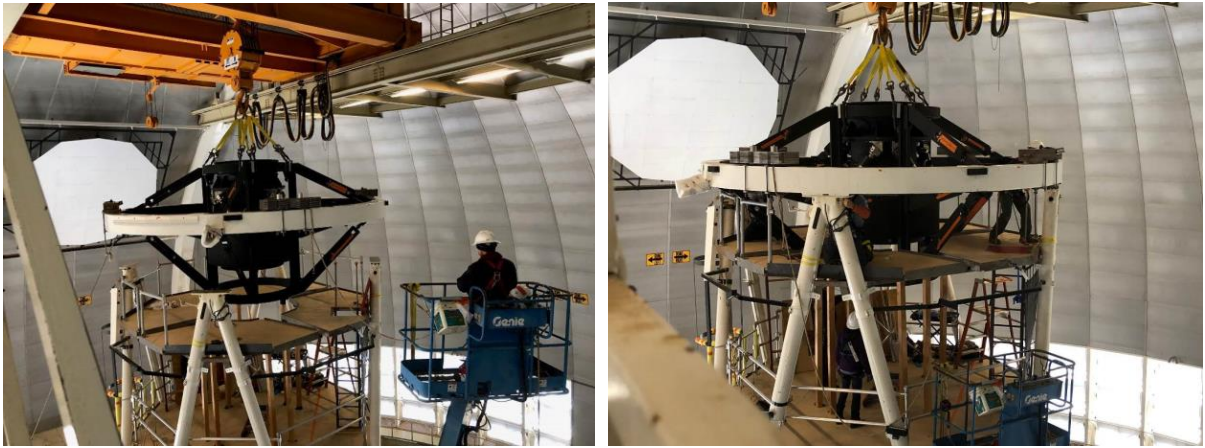

Figure 9: Left-The DESI upper assembly is lowered onto the Mayall Serrurier truss. Right-Kitt Peak personnel finalize the fastening of the ring with the truss.

The range of motion afforded the corrector by the hexapod is ±10 mm axially and ±8 mm laterally. The alignment goal was for the corrector and primary mirror to be in optical alignment with the hexapod at the middle of its range of adjustment with the telescope at zenith, at a temperature of 10-15ºC. This was accomplished by adjusting the spider vanes to position the cage and corrector such that the corrector was optically aligned with the primary mirror, with the hexapod at the middle of its range.

Measurement of the primary mirror and corrector optical locations was achieved using a laser tracker mounted to the telescope center section (Figure 10). The temporary mounting bracket positioned the laser tracker head where it had a clear line of sight to six magnetic spherical mounted retroreflector (SMR) nests permanently installed on the end of the corrector and six more bonded to the periphery of the primary mirror. The positions of the nests on the corrector had been measured relative to reference features on using a coordinate measuring machine (CMM) at Fermilab. The relation of the barrel reference features to the corrector lenses was known from the alignment process performed at UCL, linking the optical axis and focus position of the corrector system to the locations of its SMR nests.

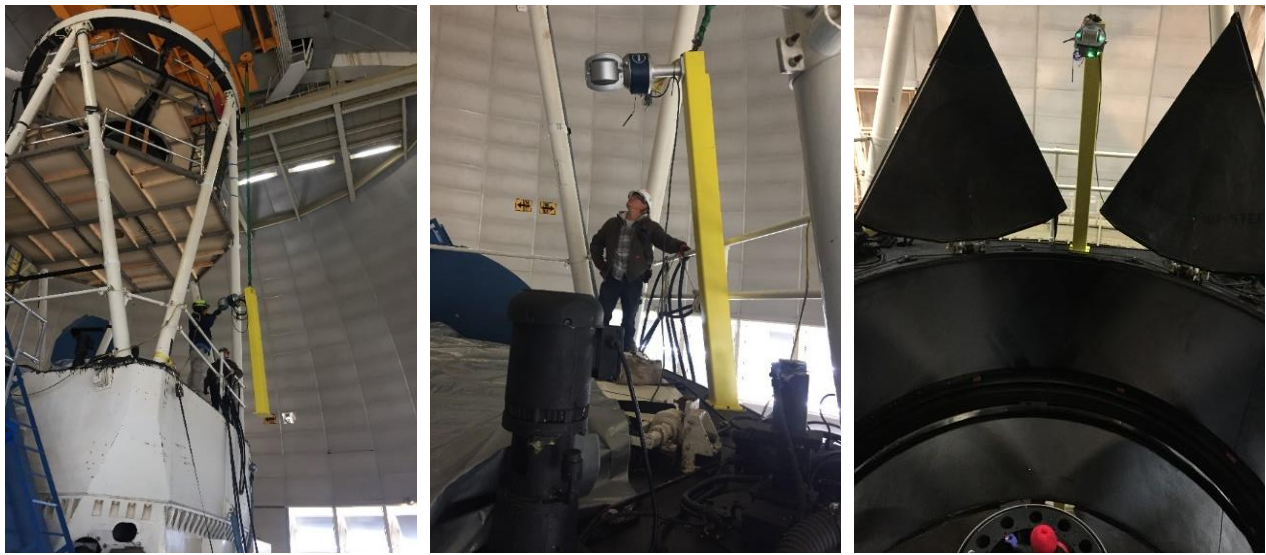

Figure 10: Left-The laser tracker on its steel mounting bracket is lifted to the telescope center section. The center panel of the floor of the lower level of the wooden work platform is removed to provide a clear line of sight to the corrector during measurements. Center-The laser tracker mounted to the top of the telescope center section, positioned so it has a clear view of the SMRs on the corrector and mirror. The mirror covers are closed and protected with plastic in this photo. Right- The laser tracker on its yellow steel mounting bracket being used to measure points through the open mirror covers. A

NOIRLab engineer in red cap at bottom of frame, positions an SMR on the telescope structure in the center of the primary mirror while the laser tracker measures its location.

The Mayall primary mirror was re-aluminized in 2014. Prior to stripping and re-coating the surface, while the mirror was removed from the telescope and resting on stands on the ground floor, six SMR nests were permanently bonded to equally-spaced positions on the outer diameter of the mirror, near the mirror surface. A laser tracker was used to measure a grid of points on the mirror surface and arcs of points in the inner bore and outer diameter of the mirror substrate, relative to the six nests. The metrology data were used to perform a best-fit of the hyperboloid mirror prescription to the surface measurements, defining the mirror optical axis and focus relative to the SMR nests and to the substrate inner and outer diameters.

This best-fit analysis of laser tracker measurements indicated that the mirror axis was offset 9 mm from the physical center of the mirror substrate, and that the mirror base radius of curvature was 18 mm smaller than the published prescription. These apparent discrepancies were the subject of much discussion within the project team. The mirror was finished in the late 1960's using the technology of that era, so these differences were plausible, but doubted by some. Based on the published accuracy of the laser tracker, the discrepancies could have been due to measurement errors, though this would have implied significant systematic bias instead of random scatter.

Initially, the laser tracker was used to align the corrector with the centerline of the mirror's physical hole and outside diameter, at the focus position indicated by the published mirror prescription (i.e., align the corrector as though the measured discrepancies were not present). This was accomplished in February 2019, before the Commissioning Instrument was installed. The state of optical alignment was later assessed on-sky by adjusting the hexapod to optimize image quality on the Commissioning Instrument. As described in Section 4.1, focusing and aligning the corrector on-sky confirmed that analysis of the laser tracker data was accurate. The mirror optical axis is offset from the mirror substrate axis, and its radius deviates from the published prescription as the laser tracker measurements and analyses showed.

## 4. FOCAL PLANE SYSTEM

### 4.1 Commissioning Instrument

The DESI Commissioning Instrument (CI), designed, assembled, and verified at OSU, was a temporarily installed imaging instrument, used as a risk-mitigation tool for the full DESI instrument[10]. It was installed and operated in the spring of 2019 to verify performance and begin commissioning of the newly-installed DESI corrector, exercising image optimization, instrument control, and data collection, with the DESI ICS commanding the telescope motion. The CI consisted of a focal plane assembly with five commercial CCD guide sensors spanning the DESI 3.2 degree diameter field of view, an enclosure for the focal plane assembly, and a set of dummy weights. These elements are shown in Figure 11.

The CI focal plane assembly mounted to the top end of the DESI corrector using the same interface features as the DESI focal plane assembly. The mass distribution of the elements of the CI were engineered to match those of the DESI focal plane system. Because the CI focal plane was simpler and lighter than the DESI focal plane, the CI enclosure served as both a protective shell and to make the total mass supported by the corrector and hexapod match the supported mass with the full DESI system installed. An additional set of dummy weights, with the same mass as the DESI Focal Plane Enclosure (FPE), was attached to the end of the cage, keeping the telescope in balance, and ensuring the upper structure behaved as it would with the full DESI system installed.

The CI succeeded in verifying the image quality of the corrector, and helped demonstrate field acquisition and guiding. In addition, it enabled on-sky, optical determination of the optimal position and orientation of the corrector relative to the Mayall primary mirror. As described in Section 3.6, the hexapod motion required to optimize the image quality in the CI confirmed the findings of the 2014 laser tracker measurements of the primary mirror. Fortunately, the range of the hexapod was sufficient to move the corrector to its optimal position. However, this position did not leave significant hexapod travel margin to account for variations in optimum due to gravity, thermal, and any other effects, so the project decided to re-align the cage so that the corrector was aligned optically with the primary mirror when the hexapod was centered. The laser tracker was again used to verify that the vane adjustments achieved the desired positioning of the cage.

The laser tracker was also used to generate a corrector deflection look-up table by pointing the telescope to various hour angle and declination combinations, measuring the location and orientation of the corrector relative to the primary mirror at each telescope pointing. This look-up table continues to be used by DESI to pre-adjust the hexapod to get close to optimal alignment for a given telescope pointing, minimizing the additional required adjustments made based on feedback from the wave-front sensing system.

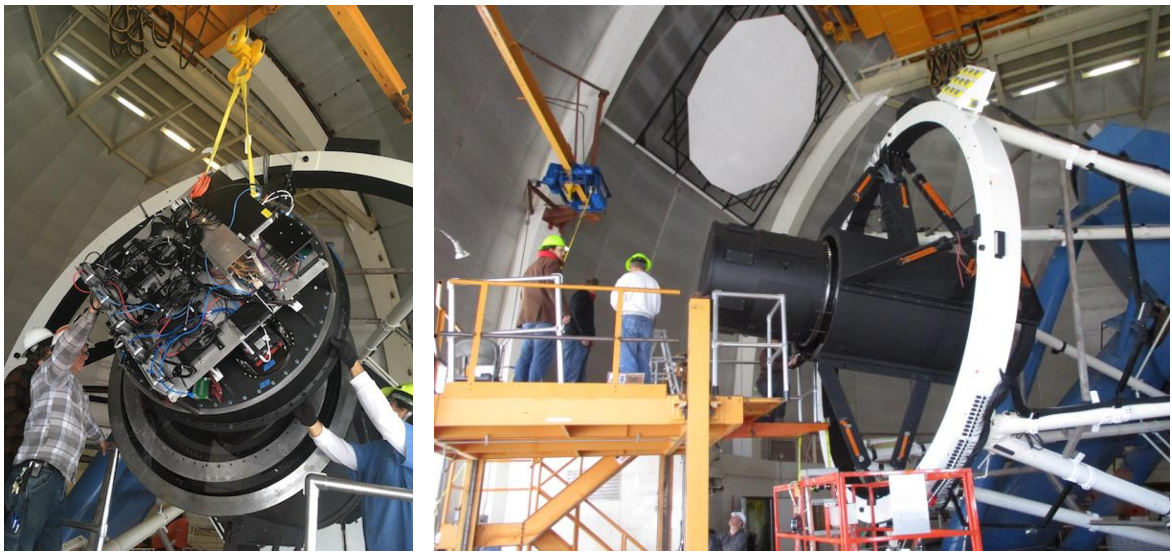

Figure 11: Left-The Commissioning Instrument focal plane assembly being positioned for installation to the DESI corrector in March 2019.  Right-The Commissioning Instrument fully installed on the telescope.

## 4.2 Mock petal installation

The DESI focal plane[11], designed and integrated at Berkeley Lab, consists of ten wedge-shaped "petals".  These are CNC-machined 36º aluminum hole plates, provided by Boston University[12].  The key components on each petal are:

- 500 robotic fiber positioners assembled at the University of Michigan[13]
- 12 fiducial illuminators from Yale University
- 1 guide/focus/alignment camera (GFA) from a consortium of four Spanish institutions: Institut de Física d'Altes Energies (IFAE), Institute of Space Sciences (IEEC-CSIC), Centro de Investigaciones Energéticas, Medioambientales y Tecnológicas (CIEMAT), and Universidad Autónoma de Madrid (UAM)[14].

The ten petals are arrayed around the optical axis, forming an 814 mm diameter focal plane.  The petal/fiber/spectrograph system is modular: a fiber cable from each petal carries the 500 optical fibers to a slithead, which is installed in one spectrograph.  The fiber cables[15] and slitheads[16] were designed and built by Durham University (United Kingdom).  The petals are mounted to the corrector via a cone-shaped adapter (the FPD, from Fermilab), which bolts to the end flange of the corrector and presents a circular mounting ring (FPR) to which the petals are attached.  At Berkeley Lab, unpopulated petals were all measured and co-aligned using shims to the FPR on a CMM.  The petals were then loaded with robots, fiducials, and GFAs, the fibers were spliced permanently to their cables, and finally the petal/fiber units were shipped individually to Kitt Peak.

To maximize the instrumented area of the focal plane, the installed petals have gaps between them of only 0.6 mm.  This necessitated very precise and linear insertion of each petal into the assembly to avoid collisions with other installed petals. Petal installation was done with the telescope stationed at the Mayall's Southeast Platform on the C floor, using a custom installation system, referred to as the "Sled", installed on the Southeast Platform.  This system is shown in Figure 12, with a dummy petal, being exercised with the focal plane simulator, described in more detail below. The base of the Sled system was a weldment that clamped to the platform.  A welded steel frame with a pair of linear rails was supported by the frame via six adjustable-length struts.  This frame was nominally angled so that the linear rails were parallel with the axis of the telescope and corrector.  A carriage with linear bearings moved along the linear rails, driven by a lead screw.  This carriage used another set of six adjustable-length struts to support the petal mount assembly (PMA), which is a steel weldment with an arm that can support a single petal on a rotary stage so that it can be inserted into any of the ten petal locations on the FPR.

A laser tracker was used for all steps of alignment of this system (see Shourt, et al., these proceedings). The six struts that support the frame with linear rails were adjusted so the direction of travel of the Sled is precisely aligned with the optical axis. The other six struts that support the PMA were then adjusted to position and align each petal to its particular target location on the FPR. Each of the six struts supporting the PMA was equipped with a load cell that indicated when the petal was properly engaged with the FPR or the petal was contacting another petal.

In April and May 2019, prior to installing the focal plane assembly to the telescope, a mock petal installation campaign was conducted on the ground floor of the telescope building, to exercise and refine the procedures for petal installation. A corrector simulator was built with an interfacing flange identical to the end of the corrector, at the same angle with respect to the ground as the corrector presents when the telescope is stationed at the Southeast Platform. The Sled assembly was placed in the same relation to the simulator as it would be when installed on the Southeast Platform. For the mock installation, dummy petals were constructed, with the same envelope, mounting features, mass, and mass center location as the populated DESI petals.

The procedures for petal alignment and installation were repeatedly exercised using this mock setup, providing invaluable lessons and helping to refine the procedures. There were aspects of the original procedures which did not work as planned, leading to changes in the procedures, with no risk to the DESI production hardware and no impact on the installation schedule.

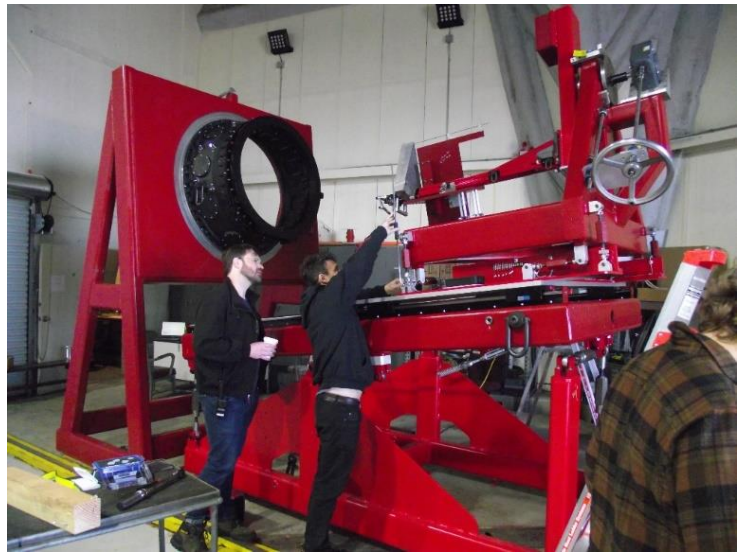

Figure 12: Berkeley Lab engineers make adjustments to the struts supporting the PMA on the Sled. The bare aluminum dummy petal is supported from the PMA. The Focal Plane Adapter (black) is already integrated to the corrector simulator weldment in the left half of the photo.

### 4.3 Integration of focal plane assembly with the corrector

During June and July 2019, the FPD and ten focal plane petals were installed to the corrector on the telescope. A portion of the flooring of the Mayall's Southeast Platform was removed to allow rigid attachment of the Sled assembly to the platform's steel crossmembers. The platform was set to a pre-determined height, then jack stands were placed under it to carry the weight load (rather than relying on the platform's existing drive system). The Sled assembly was hoisted from the ground floor, aligned relative to the telescope using the laser tracker, and attached to the platform.

Figure 13, left shows the first petal supported by the PMA and Sled, positioned for installation. Temporary cover plates protect the optics at each petal location. Figure 14, right shows a later petal, being readied for insertion next to one that has already been installed. The experience gained during the mock installation effort paid off with a smooth and efficient production petal installation campaign. The ten petals were installed, one at a time, without incident.

One unanticipated condition that the mock installation process did not simulate was gradual relative motion between the focal plane ring on the telescope and the petal on the PMA/Sled/SE Platform. In order to maximize optical stability, the telescope is mounted to a "pier" structure that is mechanically isolated from the rest of the Mayall building. The Southeast platform is fastened to the C floor, which is part of the building and isolated from the pier. When aligning the first petal with the focal plane ring, it was observed that their relative positions changed by tenths of millimeters over time periods on the order of tens of minutes. This effect is believed to be due to differential thermal effects on the external structure of the building as different areas are exposed to sunlight over the course of the day. The installation procedure was modified to minimize the elapsed time between the final adjustment of petal position and insertion into the focal plane ring. This meant, for example, that if it was close to lunch time, the final alignment and installation would be performed after lunch rather than attempted before. One or two petals were installed per working day. Figure 14 shows the front and back sides of the installed array of ten petals.

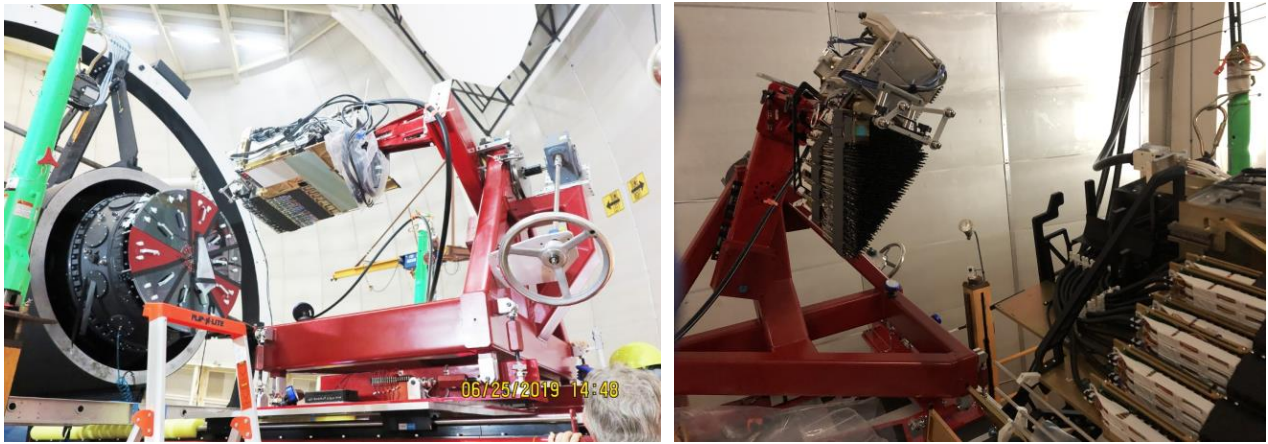

Figure 13: Left-The first of ten DESI production petals, supported on the rotary stage of the PMA, ready to be installed to the focal plane ring by advancing the Sled toward the telescope. The red and transparent wedge parts on the focal plane adapter protected the optics from the dome environment and were removed only when a petal was aligned and ready to insert. Right-A later petal ready for installation next to an already-installed petal, visible in the lower right of the photo. The bare aluminum irregular quadrilateral part above the black petal is a temporary mount for four SMRs, used to align the petal with the corrector.

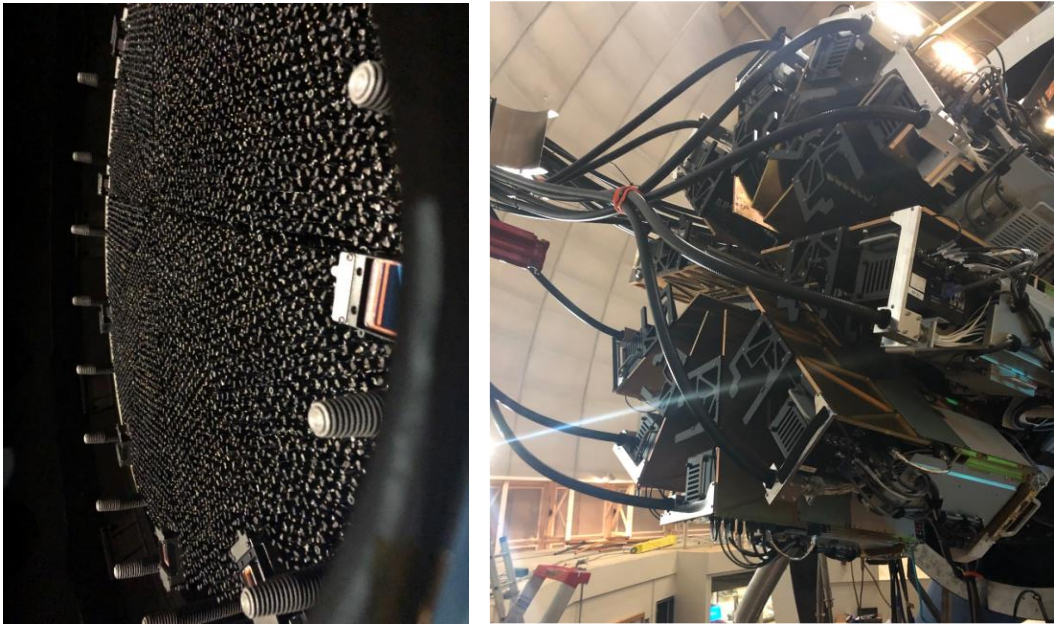

Figure 14: Left-Front side of the fully-integrated focal plane showing 5000 fiber tips on positioners, viewed through an open porthole of the focal plane adapter. A GFA is clearly visible to the right in this photo, and others are partially visible elsewhere. The ribbed posts around the periphery are black plastic covers over the shiny alignment pins on each petal,

protruding through the focal plane ring. Right-Back view of all ten focal plane petals installed to the telescope with electronics and black fiber cables visible.

### 4.4 Routing of optical fiber cables

During August and September 2019, the thermally insulated focal plane enclosure (FPE, from Berkeley Lab) was installed and the fiber cables were routed down the telescope to where the spectrographs are located. After the petals were all installed, a frame was installed around them that both supports the FPE and provides strain relief for the fiber cables aft of the petals (Figure 15, left). The frame includes four tower assemblies that extend from the top of the cage, past the petal hardware and the fiber cable strain relief. Two of these towers include gasketed pass-throughs, through which the cables emerge from the environmentally controlled inside of the FPE, helping protect the interior of the FPE from contamination, humidity, and temperature variations in the dome environment. The fiber cables go from the FPE to the upper ring, down the side of the FPE and over the Southeast and Northeast upper spider vanes, stacked on each other, to minimize blockage of light (Figure 15, right). From the upper ring, the two groups of five cables are constrained by custom hardware designed, built, and installed by NOIRLab. The two sets of cables converge into a single group of ten cables about halfway down the Serrurier truss (Figure 16, left).

To ensure the fibers allow telescope rotation about the declination and hour angle axes, without twisting the cables or violating their minimum bend radii of 200 mm, they were installed in articulated cable carriers with reversible bend directions, for the arc-shaped path the rotation requires. The bundle of ten cables coming down the telescope truss was loaded into the declination wrap carrier on a bench near the east declination bearing, which was then craned into its service position in its custom guides in the gap between the telescope center section and the hour angle "horseshoe" (Figure 16, right). Due to the limited width of this gap, the declination wrap has the cables arrayed five wide and two high in the cable carrier, with dividers between the layers of cables.

The bundle of ten cables was strain relieved to the oval tube of the horseshoe, then into the hour angle cable wrap in one layer, ten cables wide (Figure 17, right). The fixed end of the hour angle cable wrap is attached to the fixed telescope mount, and from there, the cables pass straight through penetrations in the wall of the Large Coude Room over the M floor to the east of the telescope mount. The cables continue from that wall through penetrations in the DESI spectrograph clean room shack (Figure 17, left), described in Section 5.1. The penetrations through the walls of the Large Coude Room and the shack were sized to allow the passage of the slitheads while the slitheads were in their protective shipping boxes, and insulated panels were then installed to provide environmental separation between the interior and exterior volumes.

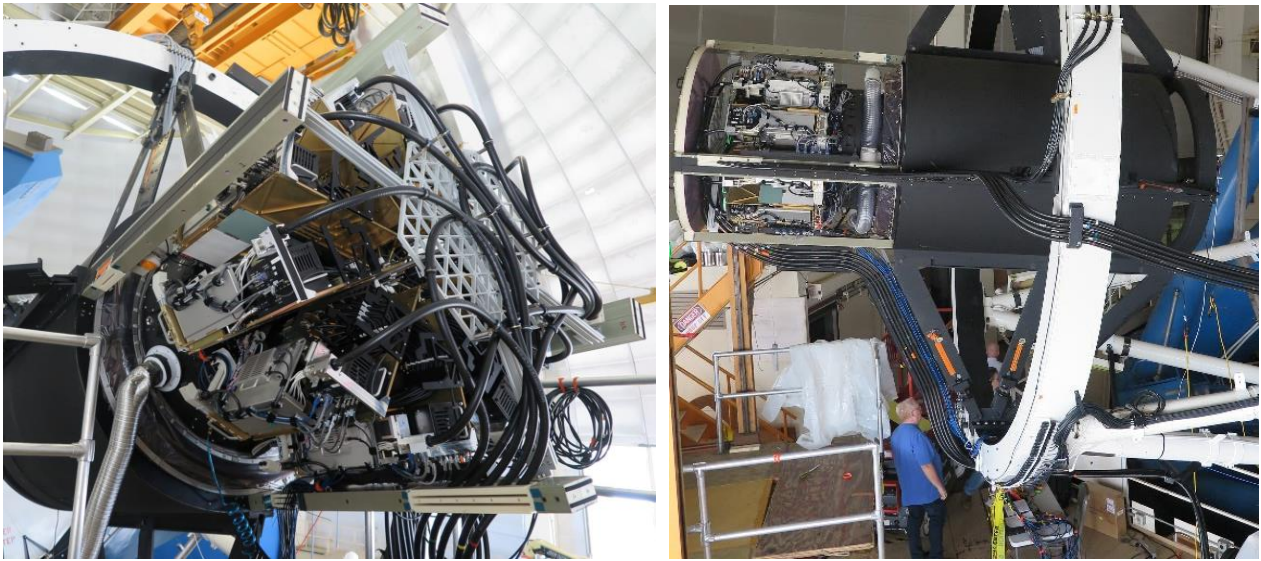

Figure 15: Left-Ten optical fiber cables from the focal plane petals are strain relieved to a frame that will later be covered within the FPE. The routing of each cable was carefully engineered to ensure the bend radius of the cables was everywhere greater than 200 mm. At the bottom of the photo, five of the cables are seen passing through the gasketed penetration of one

of the four FPE support towers.  Right-The fiber cables  penetrate the FPE volume in two groups of five, then pass over two of the upper spider vanes, over the ring, and down toward the telescope center section.

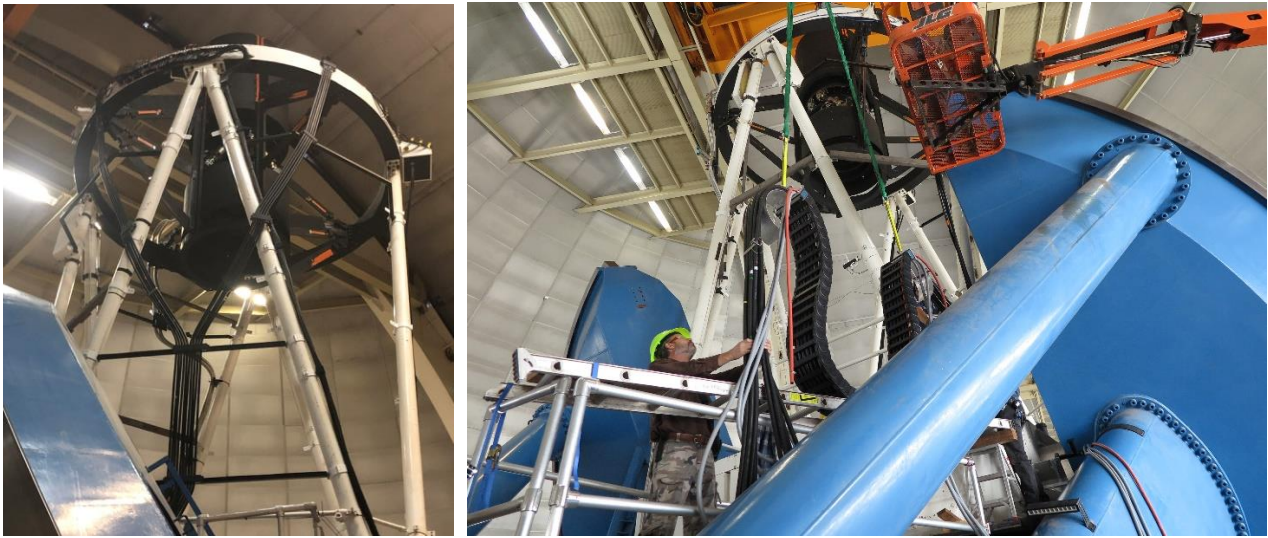

Figure 16:  Left-The two groups of five fiber cables are supported by custom mounting hardware designed and built by NOIRLab, forming a 'Y' from the upper ring down toward the telescope declination axis.  Right-Kitt Peak engineers prepare to integrate the populated articulated cable carrier around the declination axis, in the gap between the white telescope center section and the blue horseshoe.

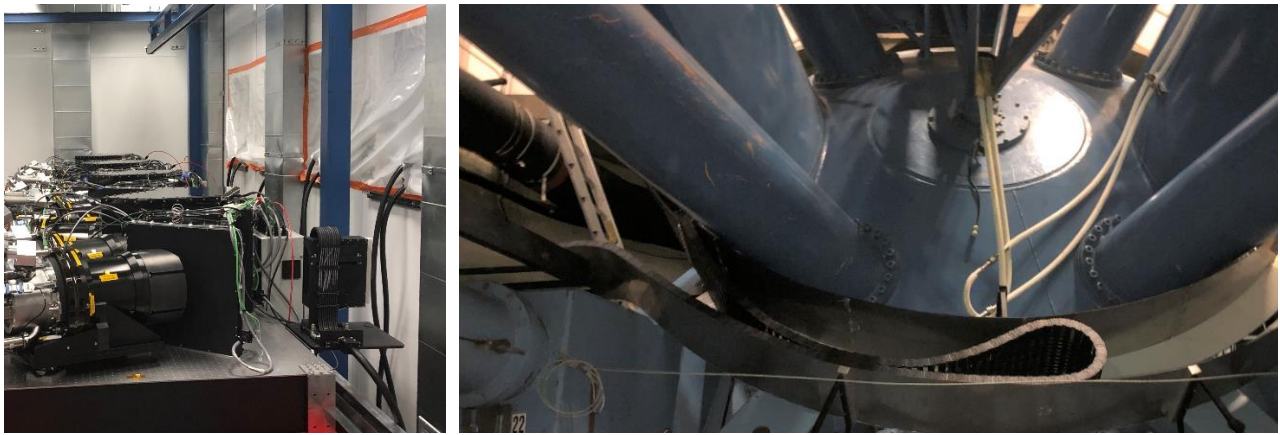

Figure 17:  Left-The fiber cables enter the spectrograph shack through temporary plastic sheet penetration covers, terminating at slitheads, mounted to shelves behind the spectrographs.  Right-The hour angle cable wrap in its sheet aluminum guides, managing the ten fiber cables between the rotating hour angle horseshoe and the fixed telescope mount.  The fiber cables run beyond the left side of this photo then through the nearby wall of the Large Coude Room on the Mayall M floor.  (The placement of these two photos roughly shows their relative positions.)

### 4.5  Fiber View Camera

To provide feedback to the fiber positioning system, the DESI Fiber View Camera (FVC, from Yale University[17]) takes images of the focal plane to measure the positions of the 5,000 optical fiber tips relative to fixed fiducial illuminators.  For these images, the optical fibers are back-illuminated by LED arrays on the exposure shutters of the spectrographs[18], so that the FVC is directly measuring the centroids of the fiber cores.  The FVC is an imaging camera with a 6k x 8k CCD sensor coupled to a singlet lens with a 600 mm focal length and 25mm aperture, producing an f/24 beam with depth of focus greater than the thermal and mechanical distance variations distance between the focal plane and the camera.  The FVC system was initially installed in 2016 for use with ProtoDESI, with some hardware upgrades occurring in 2019.

The FVC is mounted in a stiff frame that supports both the camera and the lens tube, minimizing lateral deflections as the telescope pointing changes relative to gravity. The camera and lens mounted in the frame are shown in Figure 18, left. The assembly was installed by hand to a new adapter plate which is bolted to the bottom side of the weldment that formerly supported the Mayall central baffle, at the bottom of the primary mirror cell (Figure 18, right).

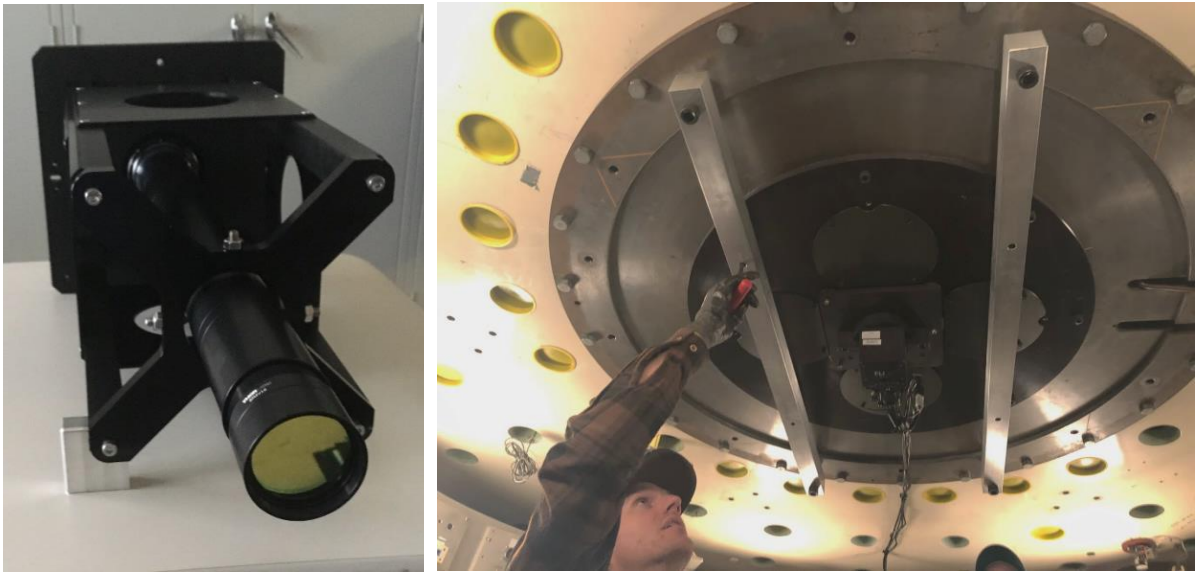

Figure 18: Left-The Fiber View Camera and its lens mounted in its frame. Right-A view from inside the Cassegrain cage of the fiber view camera (center of round features with cabling coming off of it) mounted to the round black adapter plate, which in turn is mounted to the feature on the back side of the central baffle support structure in the center of the Mayall primary mirror cell. The four "D" shaped covers around the FVC can be removed to provide arm/hand access to the lens and frame. The two bare aluminum bars support the adapter plate and FVC via ball transfers (one of which is being adjusted by a NOIRLab engineer), allowing the camera to be temporarily rotated for calibration of the system.

## 5. SPECTROGRAPH SYSTEM

### 5.1 Shack and Rack

To maintain optical alignment and minimize contamination risk, the ten DESI spectrographs require a mechanically stable, environmentally controlled space for operations that is sufficiently large to house and access them. In addition, in order to minimize light losses, the fiber cables need to be as short as possible, meaning the spectrographs need to be located as close as practical to the telescope mount. To meet the proximity and volume requirements, the Large Coude Room (LCR), directly adjacent to the telescope mount on the M floor, was chosen as the location for the spectrographs. While the walls between the LCR and the dome volume are thermally insulated, the exterior walls of the room are not well insulated, nor is the room itself equipped as a clean room. To provide thermal and environmental control, OSU designed an ISO 7 (~class 10000), thermally- and humidity-controlled clean room as large as would practically fit within the LCR, known as the shack. During the summer of 2018, NOIRLab assembled the shack, which included an internal gantry crane for installation and servicing of the spectrographs, inside the LCR (Figure 19, left).

To support the spectrographs, allow servicing access to them, and manage the cabling and fluid services to them, OSU designed and built the rack support structure, shown in Figure 19, right. The rigid structure supports the upper row of five spectrographs, while five individual carts support the lower spectrographs. The upper row is always accessible by the crane, while the lower row carts roll in and out from under the upper row for access.

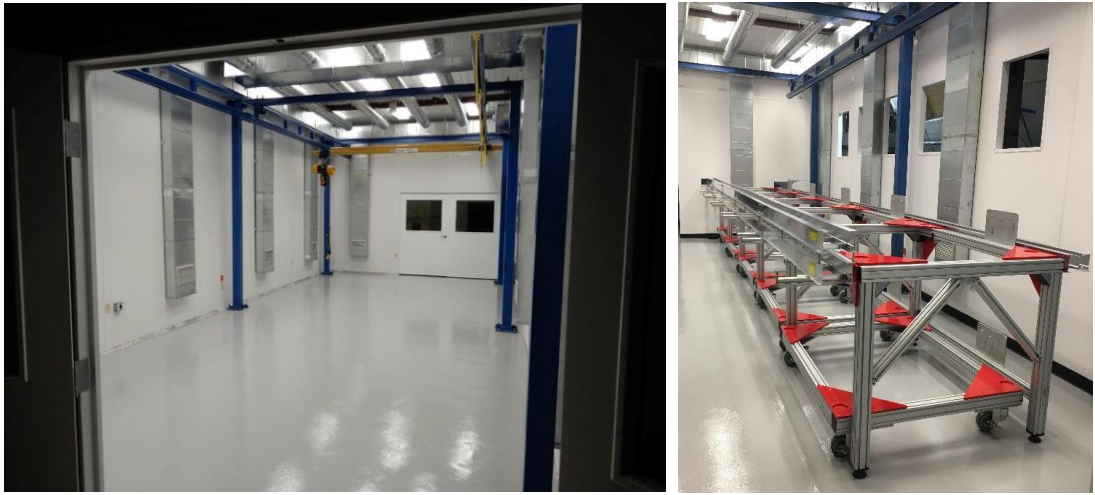

Figure 19: Left-The interior of the spectrograph clean room shack shortly after its construction and coating of the floor. The gantry crane is the yellow assembly that rides on the blue frame. Right-The spectrograph support rack in October 2018, ready for installation of the ten spectrographs in two rows of five. The upper row of spectrographs sits atop the rigid frame, while the lower row of spectrographs sit on individual carts that roll under the top row.

### 5.2 Spectrographs

Each DESI spectrograph[19] has three optical channels, each channel feeding its own CCD sensor housed in a vacuum cryostat (see Figure 20-left). Each of the thirty cryostats is a self-contained vacuum vessel, with a linear pulse tube cooler maintaining CCD cold temperatures. The cryostats[20] and their control systems were designed, built, and tested by CEA Saclay, near Paris, France. The Hartmann doors, shutters, fiber back illuminators, and mechanism controller were from OSU[21]. The complete spectrographs were integrated and tested by Aix-Marseille University[22] (AMU) in southern France, then shipped to Kitt Peak, with the cryostats each shipped separate from the main spectrograph assembly. The ten spectrographs were installed between January and December 2019.

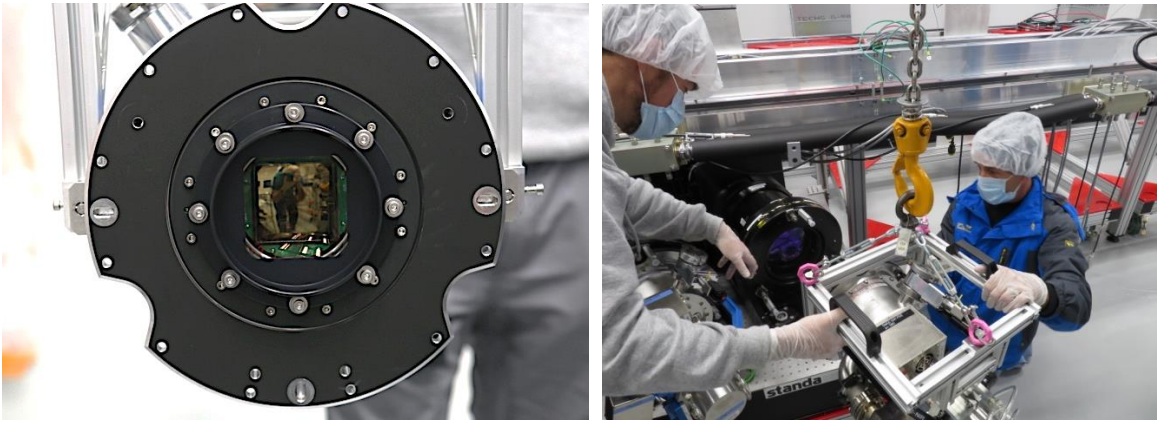

Figure 20: Left-A Blue CCD detector seen through the cryostat Field Lens. Right-Staff from CEA-Saclay and NOIRLab mounting a cryostat on a spectrograph using the dedicated lifting tool.

The spectrographs were shipped and integrated one at a time. After arrival at the Mayall, the cryostats were brought in their crates to the M floor in the personnel elevators. The spectrographs were too large for the elevators. After arrival at Kitt Peak, each spectrograph was unpacked from its shipping crate on the ground floor, then lifted to the M floor using the dome crane. The spectrograph was lowered onto a handling cart on the M floor, similar to the carts that make up the lower level of the rack, then wheeled into the LCR, then into the shack. In the shack, the cryostats were integrated to the spectrograph, using the gantry crane (see Figure 20-right). The spectrograph was lifted to either its lower-level rack cart or to its home on the fixed upper-level rack, using a custom 4-point spreader bar (Figure 21). Lower-level spectrographs were then rolled on their rack carts into position under the rigid frame of the rack. Electrical, gas, and liquid services to the spectrographs were run along the integral cable trays on the rack, then connected to the spectrographs. The slitheads at the ends of the fiber cables were integrated to their respective spectrographs, then aligned to focus the spectra on the cryostat CCDs. Figure 22 shows all ten spectrographs after their installation.

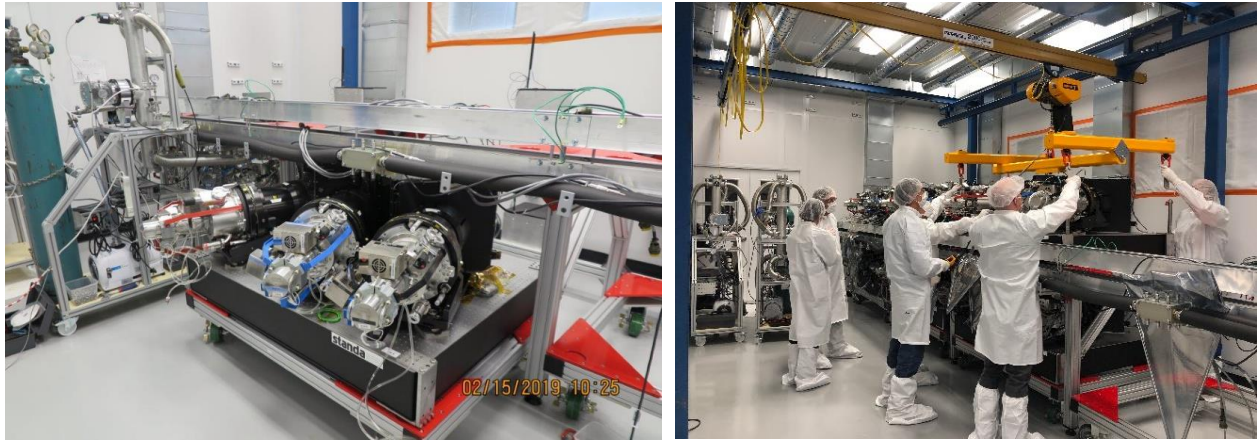

Figure 21: Left-The second installed spectrograph in the foreground, integrated with its three cryostats (shiny assemblies with red, blue, and black struts), after craning onto its lower-level rack cart. Obscured in this photograph is the first installed spectrograph, whose cryostats are being pumped down using the vacuum pump on the cart with the gas bottle on it. Right-The sixth spectrograph being lowered into its position on the top row of the rack by the shack gantry crane, using a custom spreader bar.

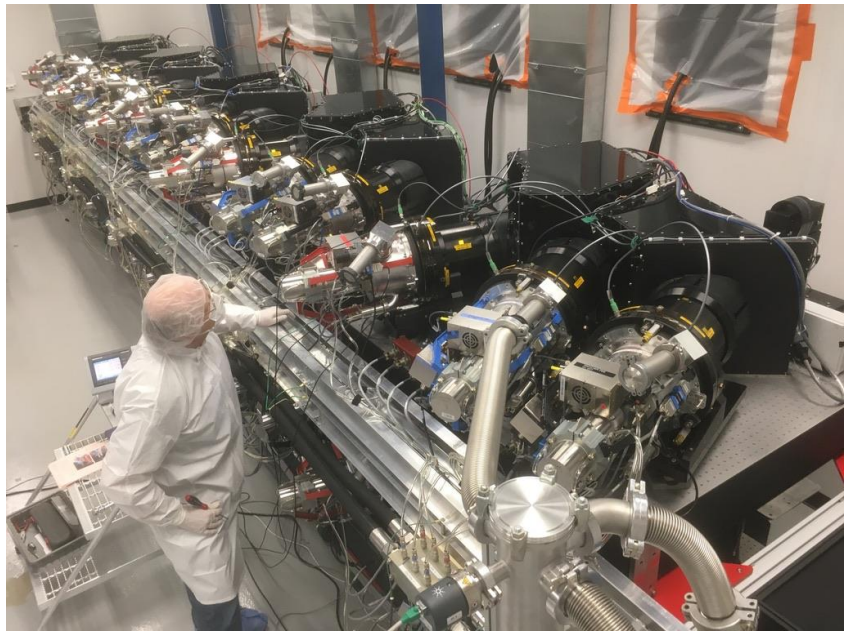

Figure 22: All ten DESI spectrographs installed and plumbed on the rack inside the clean room shack on the M floor of the Mayall building.

# 6. CONCLUSION

DESI installation at the Mayall was completed in December 2019. DESI commissioning began in October 2019 (with nine of its ten spectrographs installed) and was successfully completed in March 2020. The last tasks of the commissioning plan were performed on the night of March 15, then the instrument and telescope were powered down and put into a safe mode, with operations put on hold due to COVID-19 pandemic travel and workplace restrictions. As of November 2020, the system is being brought back online and is expected to begin survey operations in 2021.

Planning, design, construction, installation, and commissioning of DESI were successfully completed on time and under budget thanks to the hard work and commitment of all the members of the DESI team, from many institutions. DESI is a state of the art fiber-fed, multi-object, spectroscopic cosmology instrument, poised and ready to begin its five-year science survey.

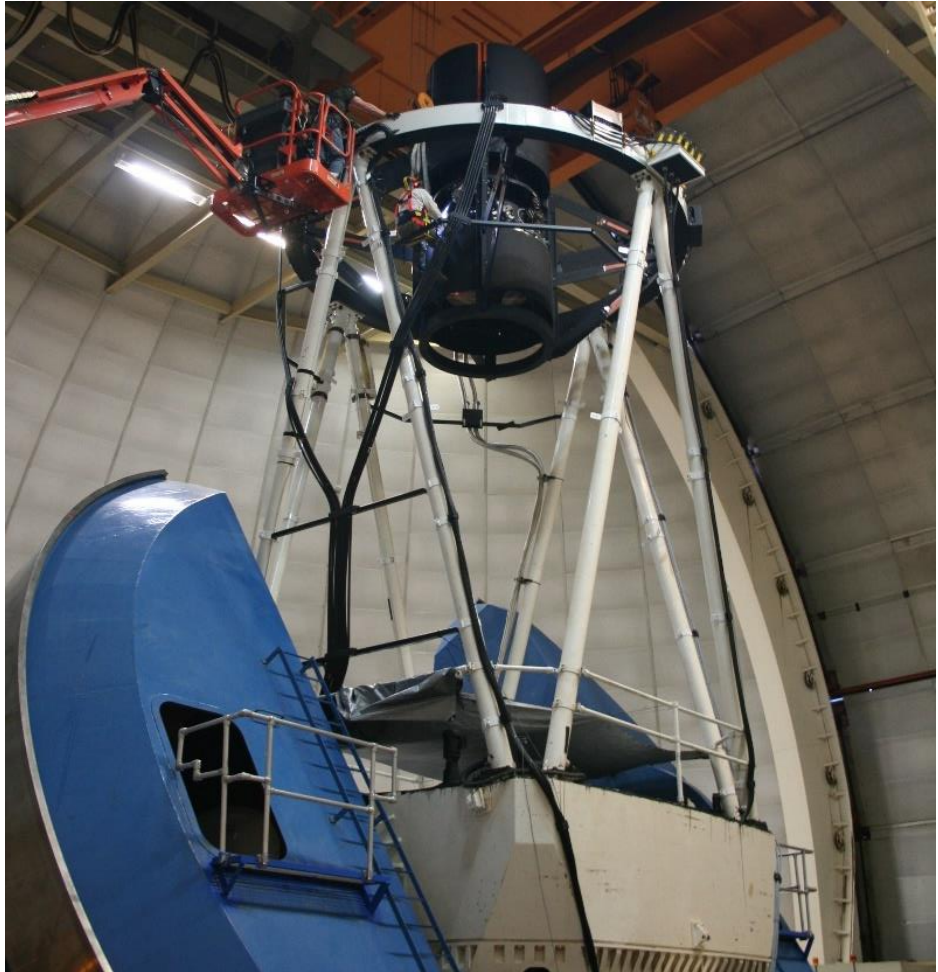

DESI and the Mayall Telescope

## ACKNOWLEDGEMENTS

This research is supported by the Director, Office of Science, Office of High Energy Physics of the U.S. Department of Energy under Contract No. DE–AC02–05CH1123, and by the National Energy Research Scientific Computing Center, a DOE Office of Science User Facility under the same contract; additional support for DESI is provided by the U.S. National Science Foundation, Division of Astronomical Sciences under Contract No. AST-0950945 to the NSF's National Optical-Infrared Astronomy Research Laboratory; the Science and Technologies Facilities Council of the United Kingdom; the Gordon and Betty Moore Foundation; the Heising-Simons Foundation; the French Alternative Energies and Atomic Energy Commission (CEA); the National Council of Science and Technology of Mexico; the Ministry of Economy of Spain, and by the DESI Member Institutions. The authors are honored to be permitted to conduct astronomical research on Iolkam Du'ag (Kitt Peak), a mountain with particular significance to the Tohono O'odham Nation.